\PassOptionsToPackage{dvipsnames}{xcolor}
\documentclass[colorlinks=true, linkcolor=blue, citecolor=blue, filecolor=blue, urlcolor=blue]{aa}

\usepackage{lipsum}
\usepackage{physics}
\usepackage{multirow}
\usepackage{xspace}
\usepackage{natbib}
\usepackage{fontawesome5}
\usepackage{wrapfig}
\usepackage[figuresright]{rotating}
\usepackage{fontawesome5}
\usepackage{listings}
\usepackage{xcolor} % For text coloring
\usepackage[varg]{txfonts}

\usepackage{float}
\usepackage{graphicx}
\usepackage{txfonts}
\usepackage{orcidlink}          % allow ORCID ids

\usepackage[hang,flushmargin]{footmisc} % remove indents in footnotes

\newcommand{\Msun}{\,M$_{\odot}$} % Solar mass
\newcommand{\MESA}{\textsc{mesa}}

\usepackage{amsmath}

\usepackage[colorinlistoftodos]{todonotes}

\begin{document}

\title{Off-centre convective zones in mass accreting stellar models}

\author{A. Miszuda \orcidlink{0000-0002-9382-2542}}

\institute{Nicolaus Copernicus Astronomical Centre, Polish Academy of Sciences, Bartycka 18, PL 00-716 Warsaw, Poland \\ \email{amiszuda@camk.edu.pl, amiszuda.astro@gmail.com}
}

\date{Received \today{}}

\abstract{
We report the physical origin of transient off-centre convective zones (oCZs) that arise in mass accreting stellar models. Using detailed MESA simulations of binary evolution, we find that these oCZs are not numerical artefacts but emerge due to a local increase in density near the retreating edge of the convective core. The density enhancement raises the local opacity, which amplifies the radiative temperature gradient $\nabla_{\rm rad}$. If this gradient surpasses the Ledoux threshold $\nabla_{\rm L}$, defined by both thermal and compositional stratification, the region becomes convectively unstable. The resulting oCZs are detached from the convective core and transient: mixing within the oCZ erases the local gradient in mean molecular weight, leaving a sharp $\nabla_{\mu}$ discontinuity at the boundary, stabilizing the adjacent layers. This mechanism naturally explains the presence and evolution of oCZs, as previously reported in massive interacting stars.}


\maketitle

\section{Introduction}

In binary stellar evolution, mass transfer episodes strongly modify the internal structure of the accretor and donor stars. This structural reconfiguration is particularly complex near the end of Roche-lobe overflow (RLOF), where convective boundaries often shift, mix, or even split. A specific feature seen in stellar evolution models after mass accretion is the presence of convective zones that are located above the convective core, disconnected from it. These so-called off-centre convective zones (oCZs) have been reported in binary models by \citet{Renzo2021}, but their physical origin remained unexplained. Subsequent studies also showed that oCZs significantly modify the internal structure of post-interaction stars \citep[e.g., see Figure 3 of][]{Wagg2024}, yet no physical mechanism has been proposed to account for their formation.
Structurally similar regions, often interpreted as semiconvective or intermediate convective zones (ICZs), have also been identified in post-main sequence massive single-star models \citep[e.g.,][]{2016ApJS..227...22F}. While the oCZs in mass gainers arise in different evolutionary and mixing contexts, their formation may be governed by similar physics involving local peaks in $\nabla_{\rm rad}$ near composition gradients.

In this Letter, we present a physical mechanism responsible for the formation of oCZs. We show that in binary models of the mass accreting star, mass accretion and enhanced mixing near the convective core boundary create a narrow region with a local increase in density.
This leads to an increase in the local opacity, which, in turn, causes a peak in the radiative temperature gradient, $\nabla_{\rm rad}$. When $\nabla_{\rm rad}$ exceeds the Ledoux temperature gradient, $\nabla_{\rm L}$, it violates the Ledoux criterion for stability and convection is temporarily triggered in a thin off-centre shell. These zones are transient, as mixing rapidly washes-out any local mean molecular weight gradients, $\nabla_{\mu}$, thereby reducing $\nabla_{\rm L}$ and restoring convective stability. Our results explain the structure and time evolution of these oCZs.

\section{Methods}
\label{methods}

We use the MESA stellar evolution code (version r23.05.1; \citealt{Paxton2011, Paxton2013, Paxton2015, Paxton2018, Paxton2019, Jermyn2023}) to compute non-rotating binary evolution models for system undergoing stable mass transfer. The models consist of a binary system with initially 10\Msun\ primary and a 7\Msun\ secondary in a circular orbit with a period of 3 days. The binary is evolved through conservative mass transfer until the accretor star reaches 10 \Msun. For subsequent evolution, we assume the accretor evolves as a single, isolated star until central hydrogen exhaustion. 

In our evolutionary computations, we adopted \cite{Asplund2009} initial chemical composition and used the OPAL opacity tables, supplemented by data from \cite{Ferguson2005} for lower temperatures. For high-temperature regimes, as well as hydrogen-poor or metal-rich conditions, we used CO enhanced tables. We assumed a metallicity of $Z~=~0.014$ \citep{Nieva2012} and an initial hydrogen abundance of $X_0~=~0.723$, using scaling relations as in \cite{2016ApJ...823..102C}.

Convective mixing is treated with the Ledoux criterion, combined with time-dependent convection following the formalism of \cite{Kuhfuss1986}, with a mixing-length parameter $\alpha_{\rm MLT} = 0.5$. Semiconvective mixing is included according to the prescription of \cite{Langer1985}, adopting $\alpha_{\rm sc} = 0.1$. Thermohaline mixing is implemented following \cite{Kippenhahn1980}, with an efficiency parameter $\alpha_{\rm th} = 1$.
Convective overshooting above the hydrogen-burning core is treated using an exponentially decaying diffusion scheme \citep{Herwig2000}, with an overshooting parameter $f_{\rm ov}=0.02$ \citep{2024ApJ...964..170J}.
As our models do not include rotational mixing, we introduce a minimum diffusive mixing coefficient of $D = 100\  \rm cm^2\,s^{-1}$ to smooth out numerical noise or discontinuities in internal profiles.

To ensure numerical convergence, we explore the numerical stability of our results against changes in both spatial and time step resolution, testing various configurations and ultimately adopting \texttt{mesh\_delta\_coeff~=~0.2} and \texttt{time\_delta\_coeff~=~0.5}.
Furthermore, to confirm the robustness of the off-centre convective zones, we recompute the models using different mixing descriptions and efficiencies. We present the results by showing comparison between the hydrogen profiles for $X_c = 0.56$, corresponding to approximately the midpoint of core rejuvenation following accretion, as shown in \autoref{appendix:mixing}. 
We note, that while the oCZs properties vary depending on the adopted treatment of mixing, these zones persist across all setups.

\section{Results}

\begin{figure}[t]
    \includegraphics[width=\linewidth]{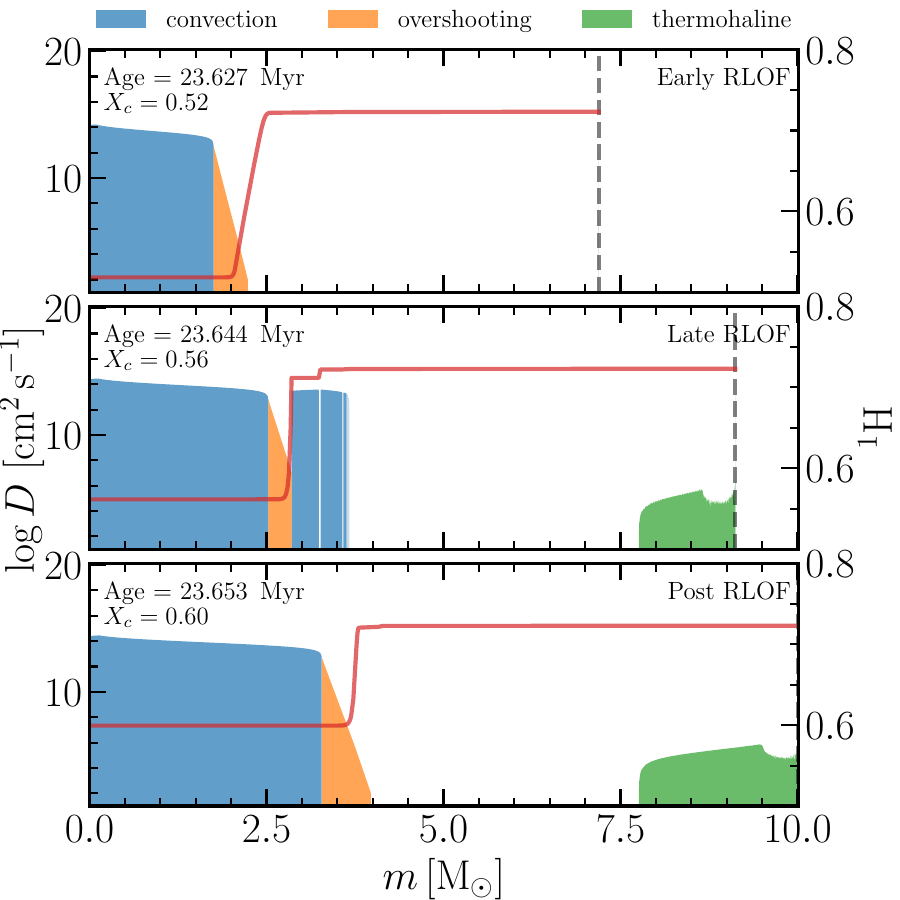}
    \caption{Internal mixing profiles and hydrogen abundance distributions (red lines) for three evolutionary stages: early, late, and post mass transfer. Coloured regions correspond to different mixing processes, including convection, overshooting, and thermohaline mixing, as labelled. The red solid lines show the internal hydrogen profiles, and the dashed vertical lines show current stellar mass.}
\label{fig:profiles}
\end{figure}

Off-centre convective zones systematically appear in our models of the mass accreting star during the late stages of RLOF. Their development is illustrated in Figure~\ref{fig:profiles}, which shows the internal mixing structure at three evolutionary stages: early, late, and post-accretion. Each panel presents diffusion coefficients associated with convection, overshooting, and thermohaline mixing. The current stellar mass of the accretor is marked by dashed vertical lines. The red curves trace the hydrogen distribution inside the star, capturing the impact of successive mixing episodes. Rejuvenation of the stellar core is evident across the sequence, indicated by the increasing central hydrogen abundance and the expansion of the convective core. 
Notably, during the late RLOF phase, multiple detached oCZs emerge. These zones are chemically homogeneous, with nearly constant hydrogen abundance, while steep hydrogen gradients appear in the narrow regions separating them.

According to the Ledoux criterion \citep{Ledoux1947}, the onset of convection depends not only on the temperature stratification but also includes stabilising effects from composition gradients:

$$\nabla_{\rm rad} > \nabla_{\rm L} \mathrm{,\ where\ } \nabla_{\rm L} = \nabla_{\rm ad} + \frac{\varphi}{\delta} \nabla_{\mu}.$$
Here, $\nabla_{\rm rad}$ is the radiative temperature gradient and $\nabla_{\rm L}$ is the Ledoux gradient composed with $\nabla_{\rm ad}$, the adiabatic gradient, and $\nabla_{\mu} = \mathrm{d} \ln \mu / \mathrm{d} \ln P$, the gradient of the mean molecular weight. The coefficients $\delta = -\left( \frac{\partial \ln \rho}{\partial \ln T} \right)_{P, \mu}$ and $\varphi = \left( \frac{\partial \ln \rho}{\partial \ln \mu} \right)_{P, T}$ are thermodynamic derivatives obtained from the equation of state \citep[e.g.,][]{Kippenhahn1990}.
When the inequality is satisfied, the stratification becomes dynamically unstable and convection is initiated, even in the presence of stabilising composition gradients.

\begin{figure}[t]
    \includegraphics[width=\linewidth]{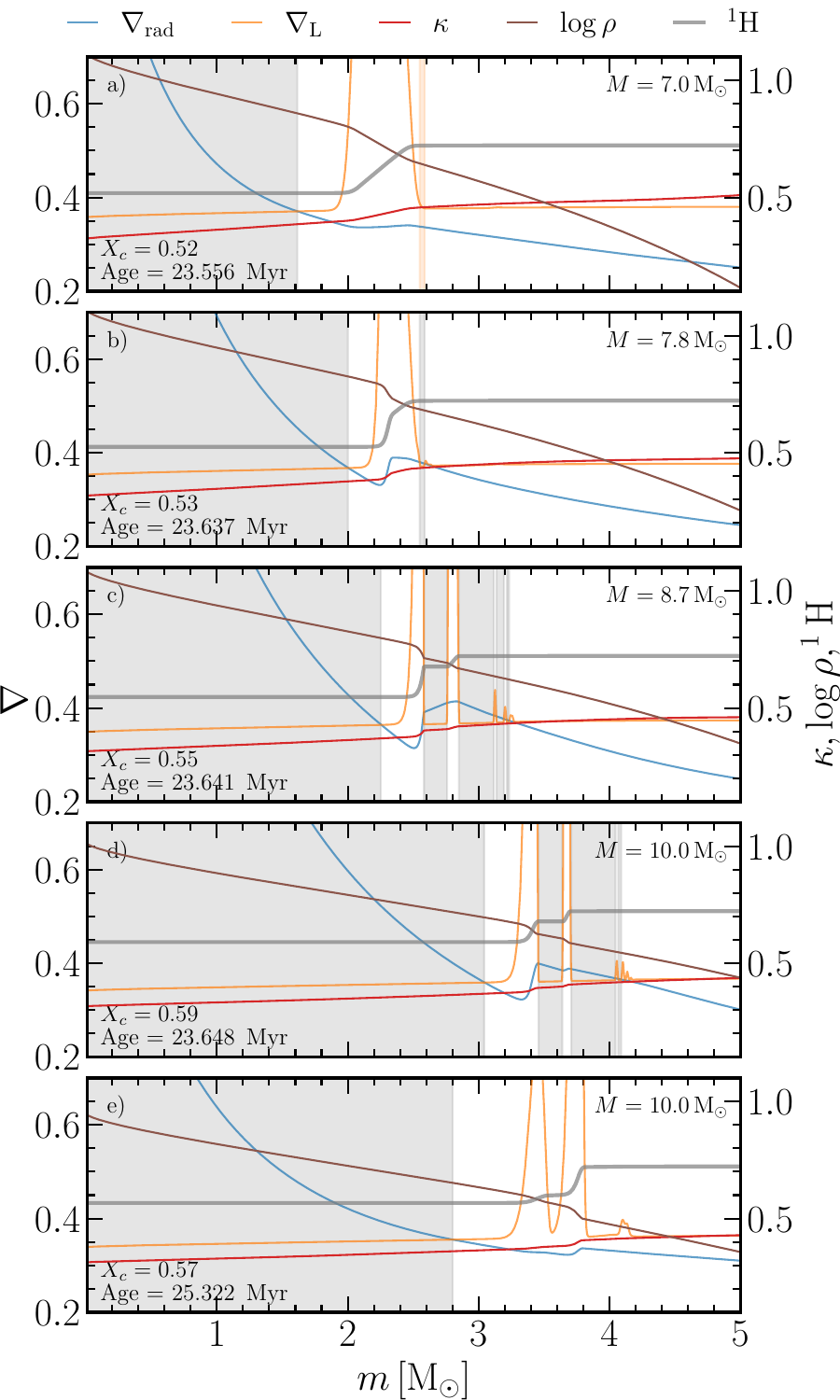}
    \caption{Temperature gradients and structural profiles in representative models before, during and after mass transfer. Shaded areas indicate convectively unstable zones.}
\label{fig:grad_kappa_rho}
\end{figure}

The internal profiles of $\nabla_{\rm rad}$ and $\nabla_{\rm L}$ for models close to those shown in Figure~\ref{fig:profiles} (compare the ages and core-hydrogen abundances between plots) are presented in Figure~\ref{fig:grad_kappa_rho}.
The right-hand ordinates display the corresponding profiles of opacity $\kappa$ (red), density $\log \rho$ (brown), and hydrogen abundance $^1$H (grey lines), plotted in the inner 5\Msun\ of the mass coordinate. The shaded regions mark the convectively unstable layers, where $\nabla_{\rm rad} > \nabla_{\rm L}$.

As the accretor grows in mass, both the density and $\nabla_{\rm rad}$ locally increase just outside the convective core (see panel b) in Figure~\ref{fig:grad_kappa_rho}, at $m\sim2.3$).
When $\nabla_{\rm rad}$ exceeds $\nabla_{\rm L}$, the Ledoux stability criterion is violated, and convection is triggered in a narrow shell located in the outermost part of the chemically stratified region, where the hydrogen gradient is still present but already very shallow. We mark this $\mu$ gradient zone (compare with the $\nabla_\mathrm{L}$ profile) in the topmost panel of Figure~\ref{fig:grad_kappa_rho}, just before the accretion started, with orange stripe for clarity. This initial convective zone, however, is quickly replaced by subsequent ones as the structure continues to evolve. The convection in these zones rapidly homogenizes the chemical composition within. However, since the surrounding layers retain $\nabla_{\mu}$ profiles, mixing leads to sharp discontinuities in composition and density at the boundaries (most visible in the panel c) of Figure~\ref{fig:grad_kappa_rho}). These discontinuities produce strong chemical gradients that locally exceed $\nabla_{\rm rad}$, thereby segmenting the oCZs. For the readers' convenience we overplot the $\nabla_{\rm rad}$ and $\log \rho$ profiles in one panel to aid tracking changes during mass accretion in Figure~\ref{fig:gradr_rho}.

The rise in $\nabla_{\rm rad}$ can be traced to structural adjustments at the core-envelope boundary following mass accretion. As fresh material is deposited onto the gainer, the convective core responds by expanding and partially mixing with the surrounding envelope. This convective boundary mixing process supplies fresh hydrogen to the core and facilitates stellar rejuvenation \citep[e.g.,][]{Neo1977,Renzo2023}. Although a strong Ledoux gradient develops in this region due to the composition discontinuity, it does not directly increase $\nabla_{\rm rad}$. However, the associated density jump and abrupt increase in hydrogen content $^1\rm H$ lead to a local enhancement in the Rosseland mean opacity, $\kappa$. This, in turn, raises the radiative temperature gradient according to
$$\nabla_{\rm rad} \propto \frac{\kappa L P}{m T^4}.$$
The increased opacity thus plays a central role in initiating off-centre convection. 
In this region, where $\log T \approx 7.3-7.4$, the opacity rise is primarily driven by bound-free and free-free transitions, which are highly sensitive to both density and hydrogen content, thereby amplifying the opacity response to structural perturbations.
Rather than the density change itself, it is the structural response to mass accretion — specifically the interplay between density and composition — that governs the local instability.

\section{Discussion}

Our models provide a self-consistent physical explanation for the emergence of off-centre convective zones in mass-accreting stars. These regions form as a result of structural changes accompanying the stellar mass growth. Analogous semiconvective layers, often referred to as intermediate convective zones (ICZs), are known to form in post-main sequence single-star models under similar structural conditions \citep[e.g.,][]{2016ApJS..227...22F}.
Specifically, the development of a narrow, higher-density layer in the region just outside the convective core leads to a local increase in opacity and, consequently, a rise in the radiative temperature gradient. 
This configuration arises naturally from the interior’s response to continued accretion and convective boundary mixing.

We confirm that the oCZs are not numerical artefacts. Their presence persists across models computed with different time resolutions and mesh settings, and their properties remain consistent. Similar structures also emerge in models with different mixing schemes and efficiencies, although their properties vary depending on the adopted treatment of mixing. These zones vanish once convective mixing within them reduces the local density gradient in the stratified region, supporting the conclusion that both their formation and disappearance are governed by physical, rather than numerical, processes.

These transient convective regions may have several implications. They modify the local $\mu$-profile, which can influence the behaviour of buoyancy \citep{Wagg2024,2024A&A...690A..65H,2025A&A...698A..49H} and pressure modes \citep[][Miszuda et al. 2025, in review]{2025arXiv250210493M} in post-interaction stars. While they do not significantly contribute to core rejuvenation, in certain configurations they may facilitate it by enhancing the transport of hydrogen into the central regions. They could also affect the timescales and spatial extent of subsequent mixing episodes. Whether similar structures form in rotating, mass accreting stars remains to be explored.

\section{Conclusions}

We identify a robust physical mechanism for the formation of off-centre convective zones in mass-accreting stellar models. These zones arise from local density enhancements that increase the Rosseland mean opacity and, in turn, the radiative gradient. Once $\nabla_{\rm rad}$ exceeds the Ledoux threshold, convection sets in within a narrow shell. Continued mixing within this region gradually smooths the local density and composition gradients, eventually suppressing the instability and restoring radiative stability.

This mechanism is repeatable and independent of numerical choices. It explains transient convective features observed in mass-accreting stellar models and highlights the importance of resolving fine-scale mixing structures. These results have implications for binary evolution, internal mixing, and the interpretation of $g$ and $p$ mode pulsations in post-interaction systems.

\begin{acknowledgements}
The author wishes to thank G. Handler and C. I. Eze for their helpful comments and suggestions that contributed to improving the manuscript.
This work was supported by the Polish National Science Centre (NCN), grant number 2021/43/B/ST9/02972.
Calculations have been carried out using resources provided by Wroc\l aw Centre for
Networking and Supercomputing (http://wcss.pl), grant no. 265. \\

Used software:
\textsc{mesa} \citep{Paxton2011,Paxton2013,Paxton2015,Paxton2018,Paxton2019,Jermyn2023},
\textsc{pyMESAreader} (\href{https://billwolf.space/py\_mesa\_reader/index.html}{https://billwolf.space/py\_mesa\_reader/index.html})

\end{acknowledgements}

\bibliographystyle{aa}
\bibliography{bibliography}

\begin{thebibliography}{25}
\expandafter\ifx\csname natexlab\endcsname\relax\def\natexlab#1{#1}\fi

\bibitem[{{Asplund} {et~al.}(2009){Asplund}, {Grevesse}, {Sauval}, \& {Scott}}]{Asplund2009}
{Asplund}, M., {Grevesse}, N., {Sauval}, A.~J., \& {Scott}, P. 2009, \araa, 47, 481

\bibitem[{{Choi} {et~al.}(2016){Choi}, {Dotter}, {Conroy}, {Cantiello}, {Paxton}, \& {Johnson}}]{2016ApJ...823..102C}
{Choi}, J., {Dotter}, A., {Conroy}, C., {et~al.} 2016, \apj, 823, 102

\bibitem[{{Farmer} {et~al.}(2016){Farmer}, {Fields}, {Petermann}, {Dessart}, {Cantiello}, {Paxton}, \& {Timmes}}]{2016ApJS..227...22F}
{Farmer}, R., {Fields}, C.~E., {Petermann}, I., {et~al.} 2016, \apjs, 227, 22

\bibitem[{{Ferguson} {et~al.}(2005){Ferguson}, {Alexander}, {Allard}, {Barman}, {Bodnarik}, {Hauschildt}, {Heffner-Wong}, \& {Tamanai}}]{Ferguson2005}
{Ferguson}, J.~W., {Alexander}, D.~R., {Allard}, F., {et~al.} 2005, \apj, 623, 585

\bibitem[{{Henneco} {et~al.}(2024){Henneco}, {Schneider}, {Hekker}, \& {Aerts}}]{2024A&A...690A..65H}
{Henneco}, J., {Schneider}, F.~R.~N., {Hekker}, S., \& {Aerts}, C. 2024, \aap, 690, A65

\bibitem[{{Henneco} {et~al.}(2025){Henneco}, {Schneider}, {Heller}, {Hekker}, \& {Aerts}}]{2025A&A...698A..49H}
{Henneco}, J., {Schneider}, F.~R.~N., {Heller}, M., {Hekker}, S., \& {Aerts}, C. 2025, \aap, 698, A49

\bibitem[{{Herwig}(2000)}]{Herwig2000}
{Herwig}, F. 2000, \aap, 360, 952

\bibitem[{{Jermyn} {et~al.}(2023){Jermyn}, {Bauer}, {Schwab}, {Farmer}, {Ball}, {Bellinger}, {Dotter}, {Joyce}, {Marchant}, {Mombarg}, {Wolf}, {Sunny Wong}, {Cinquegrana}, {Farrell}, {Smolec}, {Thoul}, {Cantiello}, {Herwig}, {Toloza}, {Bildsten}, {Townsend}, \& {Timmes}}]{Jermyn2023}
{Jermyn}, A.~S., {Bauer}, E.~B., {Schwab}, J., {et~al.} 2023, \apjs, 265, 15

\bibitem[{{Johnston} {et~al.}(2024){Johnston}, {Michielsen}, {Anders}, {Renzo}, {Cantiello}, {Marchant}, {Goldberg}, {Townsend}, {Sabhahit}, \& {Jermyn}}]{2024ApJ...964..170J}
{Johnston}, C., {Michielsen}, M., {Anders}, E.~H., {et~al.} 2024, \apj, 964, 170

\bibitem[{{Kippenhahn} {et~al.}(1980){Kippenhahn}, {Ruschenplatt}, \& {Thomas}}]{Kippenhahn1980}
{Kippenhahn}, R., {Ruschenplatt}, G., \& {Thomas}, H.~C. 1980, \aap, 91, 175

\bibitem[{{Kippenhahn} \& {Weigert}(1990)}]{Kippenhahn1990}
{Kippenhahn}, R. \& {Weigert}, A. 1990, {Stellar Structure and Evolution} ({Springer-Verlag})

\bibitem[{{Kuhfuss}(1986)}]{Kuhfuss1986}
{Kuhfuss}, R. 1986, \aap, 160, 116

\bibitem[{{Langer} {et~al.}(1985){Langer}, {El Eid}, \& {Fricke}}]{Langer1985}
{Langer}, N., {El Eid}, M.~F., \& {Fricke}, K.~J. 1985, \aap, 145, 179

\bibitem[{{Ledoux}(1947)}]{Ledoux1947}
{Ledoux}, P. 1947, \apj, 105, 305

\bibitem[{{Miszuda}(2025)}]{2025arXiv250210493M}
{Miszuda}, A. 2025, arXiv e-prints, arXiv:2502.10493

\bibitem[{{Neo} {et~al.}(1977){Neo}, {Miyaji}, {Nomoto}, \& {Sugimoto}}]{Neo1977}
{Neo}, S., {Miyaji}, S., {Nomoto}, K., \& {Sugimoto}, D. 1977, \pasj, 29, 249

\bibitem[{{Nieva} \& {Przybilla}(2012)}]{Nieva2012}
{Nieva}, M.~F. \& {Przybilla}, N. 2012, \aap, 539, A143

\bibitem[{{Paxton} {et~al.}(2011){Paxton}, {Bildsten}, {Dotter}, {Herwig}, {Lesaffre}, \& {Timmes}}]{Paxton2011}
{Paxton}, B., {Bildsten}, L., {Dotter}, A., {et~al.} 2011, \apjs, 192, 3

\bibitem[{{Paxton} {et~al.}(2013){Paxton}, {Cantiello}, {Arras}, {Bildsten}, {Brown}, {Dotter}, {Mankovich}, {Montgomery}, {Stello}, {Timmes}, \& {Townsend}}]{Paxton2013}
{Paxton}, B., {Cantiello}, M., {Arras}, P., {et~al.} 2013, \apjs, 208, 4

\bibitem[{{Paxton} {et~al.}(2015){Paxton}, {Marchant}, {Schwab}, {Bauer}, {Bildsten}, {Cantiello}, {Dessart}, {Farmer}, {Hu}, {Langer}, {Townsend}, {Townsley}, \& {Timmes}}]{Paxton2015}
{Paxton}, B., {Marchant}, P., {Schwab}, J., {et~al.} 2015, \apjs, 220, 15

\bibitem[{{Paxton} {et~al.}(2018){Paxton}, {Schwab}, {Bauer}, {Bildsten}, {Blinnikov}, {Duffell}, {Farmer}, {Goldberg}, {Marchant}, {Sorokina}, {Thoul}, {Townsend}, \& {Timmes}}]{Paxton2018}
{Paxton}, B., {Schwab}, J., {Bauer}, E.~B., {et~al.} 2018, \apjs, 234, 34

\bibitem[{{Paxton} {et~al.}(2019){Paxton}, {Smolec}, {Schwab}, {Gautschy}, {Bildsten}, {Cantiello}, {Dotter}, {Farmer}, {Goldberg}, {Jermyn}, {Kanbur}, {Marchant}, {Thoul}, {Townsend}, {Wolf}, {Zhang}, \& {Timmes}}]{Paxton2019}
{Paxton}, B., {Smolec}, R., {Schwab}, J., {et~al.} 2019, \apjs, 243, 10

\bibitem[{{Renzo} \& {G{\"o}tberg}(2021)}]{Renzo2021}
{Renzo}, M. \& {G{\"o}tberg}, Y. 2021, \apj, 923, 277

\bibitem[{{Renzo} {et~al.}(2023){Renzo}, {Zapartas}, {Justham}, {Breivik}, {Lau}, {Farmer}, {Cantiello}, \& {Metzger}}]{Renzo2023}
{Renzo}, M., {Zapartas}, E., {Justham}, S., {et~al.} 2023, \apjl, 942, L32

\bibitem[{{Wagg} {et~al.}(2024){Wagg}, {Johnston}, {Bellinger}, {Renzo}, {Townsend}, \& {de Mink}}]{Wagg2024}
{Wagg}, T., {Johnston}, C., {Bellinger}, E.~P., {et~al.} 2024, \aap, 687, A222

\end{thebibliography}

\begin{appendix}
\section{Various mixing schemes and efficiencies}
\label{appendix:mixing}

\begin{figure}[ht]
    \includegraphics[width=\linewidth]{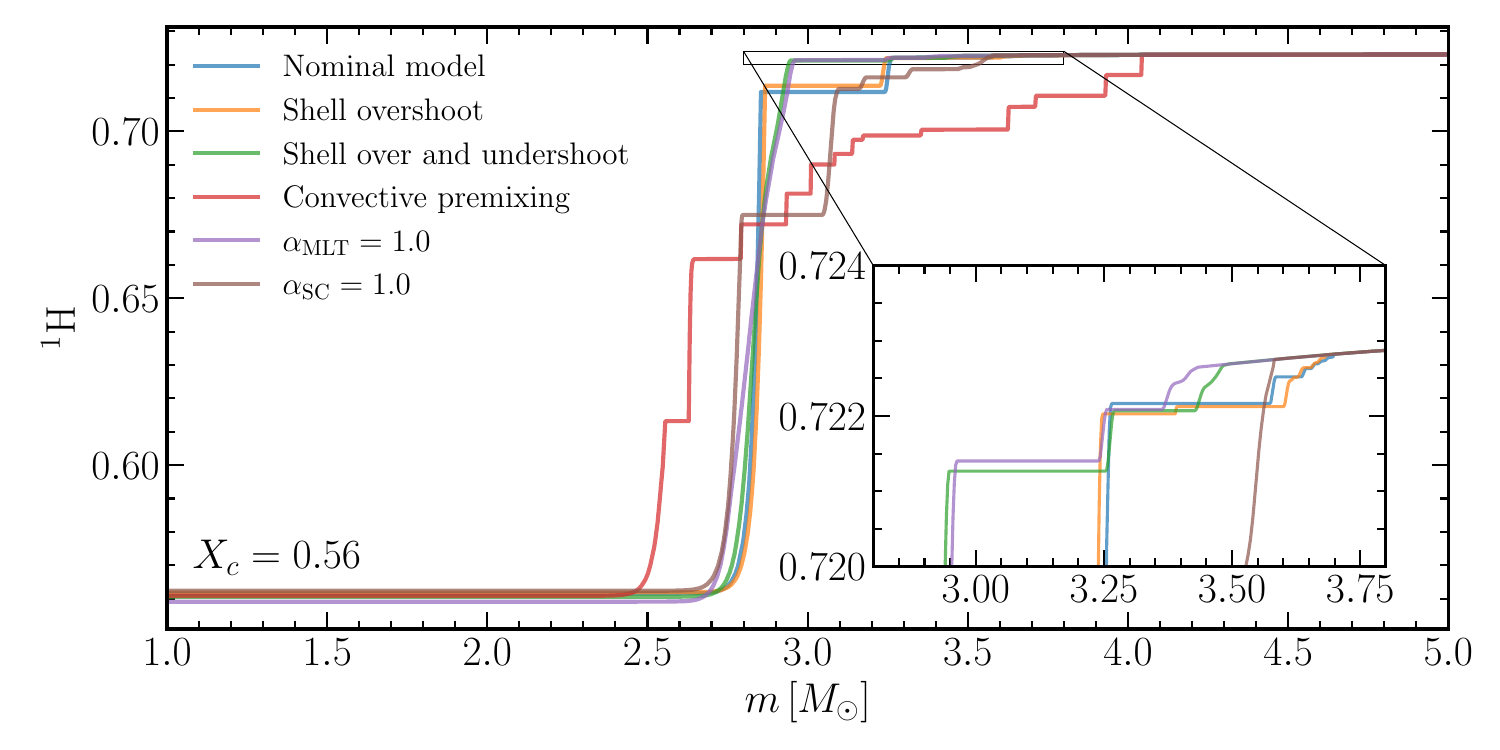}
    \caption{Internal hydrogen abundance profiles for different mixing prescriptions, taken from mass-accreting models at the evolutionary stage with central hydrogen mass fraction $X_c = 0.56$.}
\label{fig:mixing}
\end{figure}

To test the stability and robustness of the off-centre convective zones, we recompute the models using a range of mixing prescriptions and efficiencies. The goal is to assess whether oCZs consistently emerge under different assumptions. Specifically, the individual models shown here adopt: 

\begin{itemize}
    \item nominal model: model described in Section \ref{methods}
    \item shell overshoot: nominal model with exponential overshooting above the convective shells, with the efficiency of $f_{\rm ov, shell}=0.01$
    \item shell over and undershoot: nominal model with exponential overshooting above and below the convective shells, with the efficiencies of $f_{\rm ov, shell}=0.01$
    \item convective premixing: nominal model with convective premixing to set the convective boundaries (see \citealt{Paxton2019})
    \item $\alpha_\mathrm{MLT} = 1.0$: nominal model with a convective efficiency $\alpha_\mathrm{MLT} = 1.0$
    \item $\alpha_\mathrm{SC} = 1.0$: nominal model with a semiconvective efficiency $\alpha_\mathrm{SC} = 1.0$
\end{itemize}
From each model we select the mass-accreting profile with $X_c~=~0.56$ and plot the internal hydrogen distributions in Figure~\ref{fig:mixing}. All models predict the formation of off-centre convective zones, which are reflected in the characteristic step-like features in the hydrogen abundance profiles in the chemical stratification regions. The most distinct case is the model with convective premixing scheme, where the internal structure differs noticeably from the others. These differences, however, are due to the alternative algorithm used to locate convective boundaries in \MESA.

\section{Changes of \texorpdfstring{$\nabla_{\rm rad}$}{nabla	extsubscript{rad}} and \texorpdfstring{$\log \rho$}{log rho}}
\label{appendix:gradr_rho}

\begin{figure}[ht]
    \includegraphics[width=\linewidth]{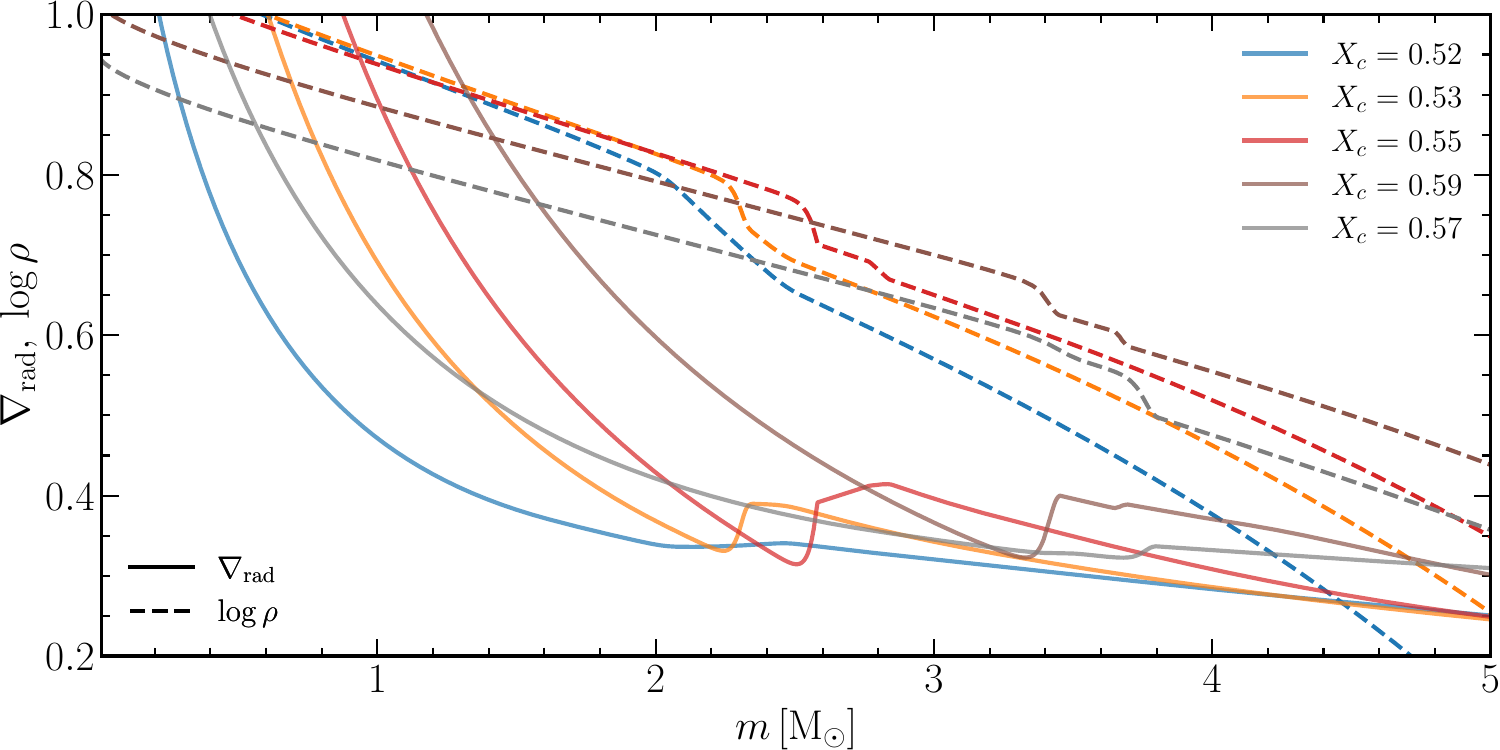}
    \caption{Changes of $\nabla_{\rm rad}$ and $\log \rho$ profiles during core rejuvenation due to mass accretion.}
\label{fig:gradr_rho}
\end{figure}

\end{appendix} 

\end{document}